\documentclass[twocolumn]{aa}
\usepackage{graphicx}

\usepackage{longtable}

\begin{document}

\title{New optical spectra and general discussion on the nature of ULX's}

\author{H. Arp
\inst{1}
\and C. M. Guti\'errez
\inst{2}
\and M. L\'opez-Corredoira
\inst{3}}

\offprints {H. Arp}

\institute{Max-Planck-Institut f\"ur Astrophysik, Karl Schwarzschild-Str.1,
  Postfach 1317, D-85741 Garching, Germany\\
   \email{arp@mpa-garching.mpg.de}
\and
Instituto de Astrof\'\i sica de Canarias, E-38200 La Laguna, Tenerife, Spain\\
\email{cgc@ll.iac.es}
\and Astronomisches Institut der Universit\"at Basel, Venusstrass 7, CH-4102 Binningen, 
Switzerland\\
\email{martinlc@astro.unibas.ch}}

\date{Received}

\abstract{
We present spectroscopic observations of three Ultra Luminous X-ray sources (ULX's). 
Two of them are very close to the active galaxy NGC~720 and the other is near NGC~1073.  
The two around NGC~720 turn out to be quasars at $z$= 0.964 and$z$= 2.216, the one near 
NGC~1073 seems to be associated to an HII region at the redshift of NGC~1073. 
We concentrate our analysis on the two
quasars and analyze them in conjunction with a set of 20 additional X-ray sources close to 
nearby galaxies which also fit the criteria of ULX's and which also have been identified as 
quasars of medium to high redshift. This sample shows an unusually large fraction of rare BL 
Lac type objects.  The high redshifts of these ULX's and 
their close proximity to their low redshift, supposedly parent galaxies is a surprising result in
the light of standard models. We describe the main properties of each of these
objects and their parent galaxy, and briefly discuss possible
interpretations.
}

\titlerunning{The nature of ULX's}

\maketitle

\section{Introduction} 

Numerous X-ray sources have been found in the central regions of nearby galaxies such as
M~31 and succesfully attributed to the presence of X-ray binary stars (e.g. Supper et
al. 1997). A number of objects with much greater luminosities in the range $L_X =
10^{39}$ to $10^{41}$  erg s$^{-1}$ have also been found in nearby galaxies. In that
case they are designated ``Ultra Luminous X-ray sources'' (ULX's).  It has been proposed
that they are exceptionally massive black hole binaries. If, however, at the distance of
the parent galaxy  a source has an X-ray luminosity exceeding about $L_X = 10^{38.5}$ erg
s$^{-1}$, it needs to be more massive than a normal stable star (Pakull \& Mirioni 2002
quote an Eddington limit as the upper bound on the luminosity of an accreting star as
$L_X = 10^{38.1}M/M_o$ erg s$^{-1}$ and Foschini et al. (2002a) quote $L_X = 10^ {38.3}$ 
ergs$^{-1}$ as the  most luminous X-ray binaries in M~31). Other explanations have been
proposed in terms of black holes in globular clusters (Angelini et al. 2001), or
associations with HII regions  (Roberts et al. 2003, Pakull \&
Mirioni 2002, Wang 2002, Gao et al. 2003). Alternative explanations have been
proposed in terms of non-isotropic emission of the radiation (King et al. 2001;
K\"ording et al. 2002). But, as discussed here, an important fraction of the
 ULX's which 
have been so far spectroscopically observed have  higher redshift than the parent galaxy. So 
currently, the nature of these objects remains unexplained, and it is necessary to rexamine 
the assumption that they are physically associated with the galaxies in which they are found.

The most extensive catalog of ULX's is the one presented by Colbert \& Ptak 2002 (CP02
hereafter) in which 87 ULX's in the near field of 54 RC3 galaxies with $cz \leq 5000$ km
s$^{-1}$ are compiled. However, these authors have filtered out known quasars
and X-ray supernova. We think this is unjustified until the true nature of the ULXs
become clear. In this paper we do not remove such objects, use only the $L_X$ criterion to
define ULX objects and change slightly the
range in X-luminosity to define ULXs,   $L_X \ge 10^{38.5}$ erg s$^{-1}$. To disentangle 
the nature of these objects it is necessary to 
find their possible optical counterparts and identify their characteristic optical spectral 
features. However, this is not an easy task  (see Foschini et al. 2002b) due to the uncertainties 
in the X-ray coordinates and the faintness of many of these objects. In this paper we present the 
optical identification  and spectroscopic observations of three ULX's, two of them (IXO1 and
IXO2 according to the notation used in CP02) in the field of the galaxy NGC~720 and one
(IXO5) around NGC~1073. 

\section{Observations and data reduction}

We found probable optical counterparts in the Digital
Sky Survey (DSS) plates for objects IXO1 and 2 around NGC~720 (the existence of 
the optical counterparts of these two objects was pointed out by CP02) and IXO5 around 
NGC~1073.  
Figure~1 shows DSS plates centered on NGC~720 and NGC~1073, and the position of
the possible optical counterparts of the IXO objects. We obtained long-slit spectroscopy of 
these three objects with the WHT\footnote{The William Herschel Telescope (WHT)
is operated by the Isaac Newton Group and the IAC in Spain's Roque de los
Muchachos Observatory.} on 28 -29 Dec. 2002.  We used the red arm of the ISIS
spectrograph with the grism R158R. The slitwidth was 1 arcsec. This provides
a sampling of 1.62 \AA\, pixel$^{-1}$ and a resolution of 8 \AA. For each
object we took a single image with a exposure time of 1800 sec. The data were
reduced using IRAF\footnote{IRAF is the Image Reduction and Analysis
Facility, written and supported by the IRAF programming group at the National
Optical Astronomy Observatories (NOAO) in Tucson, Arizona.}. Further details
of the observing conditions and data reduction can be found in
L\'opez-Corredoira \& Guti\'errez (2003).

The extracted spectra of the three objects are presented in Figure~2. All of
them are emission line objects. It is evident
from the optical spectra that objects IXO1 and IXO2 around NGC~720 are normal
appearing quasars. From the main spectral features indicated in  Figure~2, we 
obtain redshifts of $z$ =  2.216 and $z$ = 0.964 for IXO1 and IXO2
respectively. Given the paucity of quasars and the fact these have strong X-ray 
emission, we think the identification of these objects is unequivocal. 

The optical object that we have
identified  with IXO5 has strong narrow emission lines. From the position of the main lines we
measure a redshift of $z=0.0040$ or in velocity 1200 km s$^{-1}$. According to the
ratio of OIII(5007)/H$\beta$ and  H$\alpha$/NII(6584), this object should be a HII
region. Its redshift  is similar to that quoted in NED for the galaxy NGC~1073 and it is
in a region close to one of the spiral arms. The abundance of HII regions
in the spiral arms makes the identification of the optical counterpart of the X-ray
emission not as clear as for the objects IXO1 and IXO2 (furthermore there are a much
fainter optical sources near this HII region as can be seen in Figure~1). In any case,
if the identification is correct, IXO5 would be associated with ionized gas, with the X-ray 
photons of the ULX as the possible source of ionization. Gao et al. (2003) have found 
more than a dozen of ULX's apparently associated to the outer, active star-forming ring 
in the Cartwheel ring galaxy.

\section{A compilation of ULX's identified as quasars}

In Table~1 we present a compilation from the literature of objects which  fit
the definition of ULX's and have been spectroscopically identified. Three of
the objects (the ones denoted as IXO1, 2 and 43) are listed as ULX's in CP02.
The rest of the objects, although they fit the observational criteria of classification
as ULX, were not included in the above catalog apparently because they were known 
quasars with higher redshift than the parent galaxy.
The parameters quoted in this table are: 

Column 1: The designation ULX (or IXO in the CP02 catalog)
is based on a $\log L_X \geq 38.5$  erg s$^{-1}$ as estimated in Col. 5.

Column 2 and 3: The Right Ascension and declination of each object (epoch 2000).

Column 4: The X-ray counts/ks measured whenever possible from the ROSAT HRI
detector in order to match as closely as possible the CP02 flux criteria.
In a few cases outside the HRI field they were
estimated from PSPC observations.

Column 5: The X-ray luminosity $(\log L_X)$ if the ULX's are at the same
distance as the associated low-redshift galaxies (see discussion below). If
the ULX's were not at the same distance, the real luminosities would have to
be multiplied roughly by $\left(\frac{d_{\rm ULX}}{d_{\rm galaxy}}\right)^2$.
Except for the IXO's, most of these  values have been calculated by the
relation between the HRI counts/ks  and the CP02  luminosities.  

Most of the galaxies have redshift $0.003\leq z \leq 0.005$ where the
contribution to this redshift from their peculiar velocities could be
significant. Therefore, except for NGC~720, NGC~4168 and NGC~4258 the distances used for the
ULX's in Table~1 are the distance to the associated Local  Supercluster,
15.3 Mpc. (Hubble Key Project, Freedman et al. 2001). The flux calculations
vary somewhat from hard to soft bands and  the raw counts are subject to
variability in most sources. We have only  attempted to get the calculated
luminosities approximately on the scale of other  catalogs.

Column 6: The redshift ($z_Q$) measured for the object which has been 
optically identified with the X-ray source. 

Column 7: The apparent magnitude is in the V band but in a
few cases in other bands as individually noted in V\'eron-Cetti and V\'eron (2000).

Column 8 and 9: The name and redshift ($z_G$) of the host or parent galaxy.

Column 10:  The radial distance ($r$), in arcmin, from the center of the
galaxy to the  position of the ULX. This usually is about 4 - 5 arcmin for
IXO identifications but we have included a few here with slightly larger
acceptance distances as individually noted.

Column 11: The absolute optical magnitude (see above note in column 7).

Column 12: A note to indicate BL Lac type objects.  

Column 13: References

\subsection{Comments on Individual objects}
     
\begin{itemize}

\item NGC~720: These two IXO's lie very close to a bright ($B_T = 11.15$ mag) E galaxy.
The galaxy is a very strong ROSAT  X-ray source which emits 148 cts/ks. Moreover, there
is a strong, ``twisted'', X-ray  jet/filament which curves southward as it emerges from
the nucleus (Buote and Canizares 1996). In addition to the two ULX's, X-ray quasar
candidates of similar apparent  magnitude are conspicuous at larger radial separations
from NGC~720  (see ``Catalogue of Discordant Redshift Associations'', Arp 2003).

\item NGC~1073: In the CP02 catalog there are two X-ray sources in
this galaxy which are at about the same distance from the nucleus and much 
brighter in X-rays than the source which they designated IXO 5 (this is associated
to a HII region if our identification was correct, as said). These two were
not include in the IXO CP02 catalog because it was known that they were 
quasars (Arp and Sulentic 1979; Burbidge et al. 1979). But clearly these fit the 
definition of ULX's and are listed in  Table~1. Also included in Table~1 is the third 
known quasar in NGC~1073 since it has $\log L_X = 38.7$  erg s$^{-1}$ only slightly 
above the general criterion for  ULX's. The strong X-ray quasar with $z$ = 1.411 is also 
a strong radio source. Since high redshift compact objects which are strong in both 
X-ray and radio are generally now regarded as BL Lac type objects it is noted in the 
last column of Table 1. 

\item NGC~1365: This is a well known Seyfert Galaxy in the Fornax Cluster. 
About 12
arcmin to the SE in the direction of its extended spiral arm lies a very strong
X-ray BL Lac (Arp 1998, p. 49). The distance to the Fornax Super Cluster is
about the same as to the Virgo Super Cluster in the opposite direction in the
sky, or about 15 Mpc (or see Arp 2002 for a Cepheid distance of 18 Mpc). With
this  distance to NGC~1365, the BL Lac object comes within the acceptance
radius of  the IXO Catalog and is included in Table~1.

Closer in radial separation on the other side of NGC~1365 is a strong X-ray
source with $z$= 0.904 (La Franca et al. 2002). 

\item NGC~3628: This starburst/low level AGN exhibits extensive outgassed
plumes  of neutral hydrogen and is apparently ejecting X-ray material along
its minor axis. Point X-ray sources might be associated with this ejected
material and a number have already turned out to be quasars (Arp et al.
2002). We lists 7 of these  quasars which have $\log L_X$ between 38.8
and 39.6  erg s$^{-1}$.  Column 10 shows that six  of these quasars fall
within the same radial distances as most of the ULX's in  this Table. The
seventh, which has the strongest, $\log L_X = 39.6$  erg s$^{-1}$, falls at a
greater distance, very similar to the radial distance of the bright BL Lac
from NGC~1365. 

\item NGC~4151: There is a powerful X-ray source only 4.9 arcmin N of this
active Seyfert galaxy. It is also a strong radio source (de Ruiter, Arp,
Willis 1977) but, surprisingly, quite faint optically. Later it was shown
that there was an X-ray filament apparently linking this $z$= 0.615 BL Lac
quasar to the nucleus of NGC~4151 (Arp 1997, Fig. 17). The object should have
been included in the CP02 catalog but they centered their field to the SW of
NGC~4151, just leaving it out of their frame. As a result this important BL
Lac object  was not listed as an IXO. 

A very interesting object SW of NGC~4151 was, however, listed as IXO 43. It
turns out to be a galaxy on the end of a chain of seven or eight galaxies
(Arp 1977). This end galaxy on the chain has an emission spectrum of about
z = 0.239 and is apparently the identification of the Ultra Luminous X-ray
Source (IXO 43).  Three other members of the chain have redshifts near $z$=
0.160 and there are indications of absorption redshifts of $z$= 0.04 to
0.06. 

\item NGC~4168: A faint AGN galaxy at redshift $z=0.217$ 
is the ULX close to this Seyfert 1 galaxy which has $z=0.0076$ (Masetti et al. 2003). 
It is similar to the objects found in the
filament from the Seyfert NGC 7603 (L\'opez-Corredoira and Guiti\'errez 2002, 2003).
Masetti et al. title their paper "Yet another galaxy identification for an ultraluminous 
X-ray source".

\item NGC~4203: A remarkably similar object to the $z$= 0.615 BL Lac
object near NGC~4151, is the strong X-ray
source only 2.1  arcmin from NGC~4203. It has a redshift of $z$= 0.614. This
is another source which was not included in the CP02 catalog due to
previously known  high redshift.

NGC~4203 is a liner type AGN with broad Balmer lines. It is also a strong X-ray
source and an IRAS infrared source. IXO 45 is just W of the BL Lac, Tonanzintla
1480, and optically identified with a blue quasar candidate. Chandra snapshots
of this field show the galaxy X-ray emission and the strong emission of the
adjacent ULX's (Terashima and Wilson 2003).

\item NGC~4258 is a Seyfert II galaxy with "anomalous" spiral arms. The
distance used to compute the X-ray emission is 6.4 Mpc. More details can
be found in Vogler \& Pietsch 1999.

\item NGC~4319: This is an unusually disrupted spiral with a well known 
quasar/AGN, Markarian 205, only 40 arc sec south of its nucleus. As Table~1 
shows this ULX is exceptionally strong in X-rays, the strongest in the Table
at  $\log L_X = 41.5$  erg s$^{-1}$, if it were at the redshift distance of 
NGC~4319, and much larger luminosity if the redshift were cosmological.

\item NGC~4374: This is a large radio galaxy. The close X-ray source 
has long been known as a quasar of z=1.25. Again, this was not included in the CP02 catalog
just because its known high redshift.

%It can be commented as an aside that M~84 shows compression of its X-ray
%isophotes which indicates it is moving out exactly along the line of the jet
%from the radio galaxy M 87 (Arp 1987, p. 139). The z= 1.25 quasar
%therefore lies along this same line. Further out along this line from M~84 is
%the bright Sey/QSO PG 1211+143 with z=0.085. 

\item NGC~4698: The central galaxy here is a bright nearby Seyfert 2 galaxy.
Foschini et al. (2002c) identified a ULX within only 73 arcsec of the
nucleus of the galaxy. If it was an object at the same distance as the  galaxy
it would have a luminosity of $\log L_X = 39.5$  erg s$^{-1}$. This ULX, 
however, turned  out to be a quasar of $z$= 0.43. In fact it turned out to be a
pure absorption line  BL Lac quasar which is a very rare type of object.  Its 
apparent magnitude was quite red,  $B$ = 21.3 and $R$ =19.6 mag.  In order to
get  it roughly on the $V$ system we take a mean of these  and get an absolute
magnitude of $M_{abs} = -10.4$ mag, corresponding roughly to the lower limit of the
optical luminosities of the ULX/quasars in Table~1. 

{\bf Note} It is perhaps significant that almost exactly on a line through the 
nucleus of NGC~4698, extended beyond the BL Lac, is another, stronger X-ray 
source, 1RXS J124828.2+083103. It has a  redshift of $z$ = 0.120. Extended
further  beyond that is a line of fainter galaxies of unknown redshift
(Forschini et al. 2002c). It is very similar to the lines of galaxies
extending from active low  redshift galaxies that were found in (Arp 2001b).

\end{itemize}

\section{Discussion}

Data available so far and compiled in Table~1 indicates that at least an 
important fraction of the ULX's are quasars. However, there is one respect in
which the ULX quasars are different from the general run of quasars. Out of the
10 galaxies listed in Table~1, six have strong BL Lac objects associated with
them as ULX's.  A further case was recently
discovered: in a recent Chandra investigation of the multiple interacting
nuclei of NGC~4410 a possible population of ULX sources were investigated.
Nowak et al. (2003) reported: ``One of these sources is clearly associated with
a radio point source, whereas another source (in fact the brightest in the
system), may also be associated with radio emission''.

\subsection{Excess of X-ray quasars around active galaxies}

A first observation is that almost every one of the galaxies which have 
associated ULX's in Table~1 are Seyfert, starburst, radio or AGN active
galaxies. One might argue that these are selectively the most X-ray observed
class of galaxies, nevertheless it is striking that there is not even one,
normal quiescent galaxy among the group. A systematic study of the kinds of
galaxies most associated with quasars and ULX's would be useful. Apparently this
is in disagreement with the paper by Humphrey et al. (2003) which
states that they found, with Chandra, 22 ULX's in a sample of 13 {\it normal}
galaxies. However the 7 galaxies  from which those 22 ULX's actually
come,  include a  Liner/AGN, a Seyfert, an X-ray
cluster, an interacting radio X-ray pair (Arp 281) and a strong radio ejecting
galaxy in the center of the Fornax Cluster (NGC~1399). For their five least
active galaxies only one candidate ULX was found. So, it seems that in fact 
the active galaxies do preferentially produce the ULX's.

In their study of X-ray sources in fields around bright Seyfert galaxies,
Radecke (1997) and Arp (1997) showed that there was an overdensity of
bright, point X-ray sources compared to control fields at a significance
level of  $7.5\sigma$. According to the present paper, these X-ray sources
consistently turn out to be mostly quasars of a normal range in redshift.
The radii from the central galaxy within which these X-ray quasars were
found was roughly 5 to 50 arcmin (from the outer edges of the galaxies to
the field limit of the PSPC, ROSAT detector). In the present case of the
ULX's, the majority are found within about 5 arcmin of the central galaxy.
Roughly speaking, there is a difference of about 100 in the areas concerned
so that the chance of finding unrelated background sources is roughly 1/100
of the probability in the larger area in which the $7.5\sigma$ association
was found. When some ULX's turned out to be much higher redshift, each of these
was relegated, on an individual basis, to be a background source. Now,
however, we have about 20 cases, all turning out to be quasars but not one
with a binary stellar spectrum.

Of the 24 Seyferts investigated in Radecke (1997) and Arp (1997) only four
had BL Lac's associated. But it should be noted that those BL Lac's were
the most significant associations with the surveyed active galaxies. Just
by themselves the chance of finding those four was $2\times 10^{-9}$. Here
in Table~1 we see two of those previous BL Lacs (with  NGC~1365 and
NGC~4151) but in addition, four new ones, four of which are
extremely close, between 2.1 and 1.2 arcmin of the host galaxy so the
probability to have them as background sources is much lower. The question
then is: why would BL Lacs be more closely associated than ordinary quasars
with Seyfert, starburst, radio or AGN galaxies?

\subsection{Very bright, off nuclear, Chandra ULX's}

Active galaxies like NGC~3628, Arp~220 and M~82 are now known to have very
bright, ULX sources very close to their nuclei. Because these are generally
even more X-ray luminous than the ones discussed here it would be
especially important to attempt optical identifications with high
resolution telescopes. For instance in NGC~3628 a source $\log L_X = 40$
only 20 arcsec from the nucleus is reported  as violently variable
(Strickland et al. 2001). Arp 220 has a source $\log L_X = 40.6$  erg
s$^{-1}$ only 7 arcsec off  nucleus (Clements et al. 2002). NGC~7319, the
Seyfert Galaxy in Stephan's Quintet has a ULX only 8 arcsec S of its
nucleus.(Trinchieri et al. 2003); it is identified with a stellar 
appearing object at $B$ = 21.78 mag (Galianni \& Arp in preparation). M~82
has its most luminous X-ray source variable between $\log L_X \sim 40$ to
41  erg s$^{-1}$ (Matsumoto et al. 2001) lying within 8 arcsec of the  M82
nucleus, but no optical identifications have been made so it can only be
surmised that it would  turn out to be like, or a precursor of, the ULX's
for which we have spectra in the present paper. 

%Possibly the strongest ULX known it is only 40 arcsec south of the
%disrupted  galaxy NGC~4319. If Markarian 205 were at the  same distance as
%NGC~4319 (both objects are apparently joined  by a low surface brightness
%bridge within which is a narrow connection as can be seen on ground based
%telescopes, see Arp 1987; 2003; or on HST, see Schilling 2002) its X-ray
%luminosity would be $\log L_X = 41.5$  erg s$^{-1}$. Its absolute 
%magnitude  would be $M_{abs} = - 16.5$ mag. 

\subsection{Possible interpretation}

Arp (1987,1998) suggested that quasars emerge from the nuclei of 
galaxies.  In that model the quasars or proto quasars must pass out from
the nucleus and out through the body and immediate environments of the
galaxy before arriving at the distances where they have been shown to be
physically associated. In this case they must pass through the region in
which the present ULX's have been identified. It was therefore {\it
predicted} that we should observe quasar-like objects in just the region in
which observers are reporting ULX's (see Burbidge, Burbidge and Arp 2003a).
Seyfert, starburst, radio or AGN active galaxies  would be the parent
ejecting galaxies.  Rare cases such as the ULX in NGC~4698 (see Table~1),
where there are no optical emission lines would represent a recent or
strong internal burst or external  interaction which removed even the usual
gaseous, optical emission regions  leaving only the core of absorption line
stars. This model also gives a possible evolutionary connection between
recent observations of HII and other narrow line galaxies and the ejection
of quasars.  Several high or medium redshift emission line objects
apparently  emerging from low redshift active galaxies either in or
entraining luminous filament/arms  are given in the literature: NGC~4151
(Arp 1977, 2003), NGC~1232 (Arp 1982, 2003), NGC~7603 (L\'opez-Corredoira
\& Guti\'errez 2002, 2003), NEQ3 (Arp 1977, Guti\'errez \&
L\'opez-Corredoira 2003), NGC~1365 (Roy \& Walsch 1997), NGC~3628 (Arp et
al. 2002), 53W-003 (Arp 1999b), NGC~2639 (Arp, Burbidge and Zibetti, A\&A
submitted), etc. Many of these systems are known to have ULX's. Obviously 
deeper multi-color images and Chandra\footnote{It must be 
noted, however, that an application to make an observation of NGC~7603
with  Chandra was turned down; as a consequence we do not know if the two
HII  galaxies in the bridge are X-ray sources nor do we know what else this
active  Seyfert contains.} X-ray observations should yield important
results on all  of these objects. 

Another suggested explanation for these association of galaxies and quasars with
different redshift is mesolensing by King objects. In L\'opez-Corredoira \&
Guti\'errez (2003) this is discussed in more detail.

\section{References}
\noindent

Angelini, L. et al. 2001. ApJ, 557, L35

Arp, H. 1977, ApJ, 218, 72

Arp, H. 1982, ApJ, 263, 54

Arp, H. 1987, Quasars, Redshifts and Controversies 1987, Interstellar Media,
Berkeley

Arp, H. 1997, A\&A, 319, 33

Arp, H.C. 1998, "Seeing Red"(Apeiron, Montreal)

Arp, H. 1999b, ApJ, 525, 594

Arp, H. 2001b, ApJ, 549, 802

Arp, H. 2002, ApJ, 571, 615

Arp, H. 2003, "A Catalogue of discordant Redshift Associations", Apeiron,
Montreal, in press

Arp, H., \& Sulentic, J. 1979, ApJ, 229, 496

Arp, H., Burbidge, E. M., Chu, Y., Flesch, E., Patat, F., \& Rupprecht, G. 
2002, A\&A, 391, 833

%Arp, H.C., Bi, H.G., Chu, Y., \& Zhu, X., 1990 A\&A, 239, 33

Buote, D., \& Canizares, C. 1996, ApJ 468, 184

Burbidge, E.M., Junkkarinen, V., \& Koski, A. 1979, ApJ 233, 97

Burbidge, E.M., Burbidge, G.R., \& Arp H.C. 2003, A\&A, 400, L17

%Burbidge, E.M., Burbidge, G.R., Arp H.C., \& Zibetti, S. 2003b, ApJ, in press

Clements, D. et al. 2002, ApJ, 581, 974

Colbert, E. \& Ptak, A. 2002, ApJS, 143, 25

de Ruiter, H., Arp, H., \& Willis, A. 1977, A\&AS, 28, 211

Foschini, L., Di Cocco, G., Ho, L. C., et al. 2002a, A\&A, 392, 817

Foschini, L., et al. 2002b, astro-ph/0209298

Foschini, L., Ho, L., Masetti, N., et al. 2002c, A\&A, 396, 787

Freedman, W., Madore, B., Gibson, B. et al. 2001, ApJ, 553, 47.

Gao, Y., Wang, Q. D., Appleton, P. N., \& Lucas, R. A. astro-ph/0309253

Guti\'errez, C. M, \& L\'opez-Corredoira, M. 2003, A\&A, submitted (letter)

Humphrey, P., Fabbiano, G., Elvis, M., Church, M., Balucinska-Church, M. 2003,
astro-ph/0305345

King, A. R., Davies, M. B., Ward, M. J., Fabbiano, G., \& Elvis, M. 2001, ApJ,
552, L109

K\"ording, E., Falcke, H., \& Markoff, S. 2002, A\&A, 382, L13

La Franca, F.; Fiore, F.; Vignali, C., Antonelli, A., Comastri, A., Giommi, P., 
Matt, G., Molendi, S., Perola, G., Pompilio, F. 2002, ApJ 570, 100

L\'opez-Corredoira, M., \& Guti\'errez, C. M. 2002, A\&A, 390, L15

L\'opez-Corredoira, M., \& Guti\'errez, C. M. 2003, A\&A, submitted

Masetti, N. et al. 2003, ApJ, 406, L27

Matsumoto, H., Tsuru, T., \& Koyama, K., 2001, ApJ, 547, L25

Nowak, M., Smith, B., Donahue, M., \& Stocke, J. 2003, AAS Head Meeting 35, 13.16

Pakull, M.W. and Mirioni, L. 2002, astro-ph/0202488

Radecke, H.-D. 1997, A\&A 319, 33

Roberts, T. P., Goad, M. R., \& Warwick, R. S. 2003, MNRAS, 342, 709

Roy, J.-R., \& Walsh, J. 1997, MNRAS, 288, 715

%Schilling, G. 2002, {\it Science}, 298, 345

Strickland, D., Colbert, E., Heckman, T., et al. 2001, ApJ, 560, 707

Supper, R., Hasinger, G., Pietsch, W., et al. 1997, A\&A, 317, 328

Terashima, Y., \& Wilson, A. 2003, ApJ, 583, 145

Trinchieri, G., Sulentic, J., Breitschwerdt, D., Pietsch, W. 2003, A\&A, 402, 73

V\'eron-Cetty, M.P. and V\'eron P. 2000, A Catalogue
of Quasars and Active Nuclei, 11th Edition. A\&A 2003 (in press)

Vogler, A. \& Pietsch, W. 1999, A\&A 352, 64
 
Wang, Q. D. 2002, MNRAS, 332, 764

\fontsize{3}{4}\selectfont
\onecolumn
\scriptsize
%\begin{table}[thp]
%\begin{center}
%\caption{Properties of objects analyzed in this paper}
\begin{tabular}{lccccccccclll}
\fontsize{3}{4}\selectfont
\\

ID & RA  & Dec & cts/ks & log (L$_X$) &$z_Q$& mag & Host galaxy & Z$z_G$ & r &
M$_{\rm abs}$ & Type & Refs.\\
 & (hh mm ss) & (deg mm ss) & & & & & & &  (arcmin)\\
\hline
\\
IXO1& 01  52 49.7 &  -13 42 11 &	0.50   & 39.1	     &    2.216 &  20.6 	&   NGC~720 &    0.006 &   3.4 
& -11.3       & ...  & 1\\
IXO2& 01  53 02.9 &  -13 46 53 &	1.77   & 39.7	     &    0.964 &  19.2 	&   ...     &    ...   &   2.7  & -12.7       &  ... & 1 \\
\\
ULX&  02  43 39.4 &   01 21 09 &	1.2    & 39.2	     &    1.400 &  20.0 	&   NGC~1073&    0.004 &   1.4  &  -10.9      & BL Lac & 2\\
ULX&  02  43 39.5 &   01 22 20 &	0.4    & 38.7	     &    1.945 &  19.8 	&   ...     &    ...   &   1.8  &  -11.1      &...    & 2\\
ULX&  02  43 33.3 &   01 21 36 &	3.0    & 39.6	     &    0.599 &  18.8 	&   ...     &    ...   &   2.1  &  -12.1      &  ... & 2 \\
\\
ULX&  03  34 07.1 &  -36 04 00 &	3.6    & 39.8	     &    0.904 &  19.7 	&   NGC~1365&    0.005 &   7.7  &  -11.2      & ... & 3\\
   &  03  33 12.2 &  -36 19 48 &	38.8   & 40.9	     &    0.308 &  18.0 	&   ...     &    ...   &  12.4  &  -12.9      &  BL Lac & 3\\
\\
%ULX& 06  13 43.3 &   71 07 27 &      9.5-12.9 & 41.0-41.6   &    0.267 & 18.5, 19.6	&   MARK 3  &    0.014 &  10.5  & -14.1, -15.2&  BL  Lac \\
%\\
ULX&  11  20 14.7 &   13 32 28 &	0.83   & 38.8	     &    0.995 &  20.1 	&   NGC~3628&    0.003 &   3.0  &  -10.8      & ... & 4\\
   &  11  20 11.9 &   13 31 23 &	1.39   & 39.0	     &    2.15  &  19.5 	&   ...     &    ...   &   4.2  &  -11.4      & ... & 4\\
   &  11  20 06.2 &   13 40 24 &	2.50   & 39.3	     &    0.981 &  19.2 	&   ...     &    ...   &   5.5  &  -11.7      & ... & 4\\
   &  11  20 39.9 &   13 36 20 &	0.83   & 38.8	     &    0.408 &  19.6 	&   ...     &    ...   &   5.6  &  -11.3      & ... & 4\\
   &  11  20 41.6 &   13 35 51 &	0.83   & 38.8	     &    2.43  &  19.9 	&   ...     &    ...   &   6.0  &  -11.0      & ... & 4\\
   &  11  19 46.9 &   13 37 59 &	1.67   & 39.1	     &    2.06  &  19.6 	&   ...     &    ...   &   7.9  &  -11.3      & ... & 4\\
   &  11  21 06.1 &   13 38 25 &	6.11   & 39.6	     &    1.94  &  18.3 	&   ...     &    ...   &  12.3  &  -12.6      & ... & 4\\
\\
IXO43&12  10 07.9 &   39 23 12 &	0.88   & 39.3	     &    0.239 &  18.4 	&   NGC~4151&    0.003 &   5.0  &  -12.5      &... & 5\\
  &   12  10 26.7 &   39 29 09 &	63-106 & 40.6-41.9   &    0.615 &  20.3 	&   ...     &    ...   &   4.9  &  -10.6      &  BL Lac &5\\
\\
ULX&  12  12 14.6 &   13 12 48 &         4.0   & 39.4        &    0.217 & 18.7          &   NGC~4168&    0.0076&   0.75
&  -13.9      &... & 6\\
\\
ULX&  12  15 09.2 &   33 09 55 &	38.4   & 40.7	     &    0.614 &  17.5 	&   NGC~4203&    0.004 &   2.1  &  -13.4      & BL Lac & 7\\
\\
ULX & 12  19 23.6 &  +47 09 38 &         3.9   & 38.9        &    0.398 &  20.4         &  NGC~4258&   0.0015& 8.6
&-8.6 &...  & 8 \\
  \\
ULX&  12  21 44.0 &   75 18 38 &        255    & 41.5        &    0.071 &  15.2         &   NGC~4319&
   0.005 &   0.7  &  -15.7     &BL Lac & 9 \\
\\
ULX&  12  25 12.0 &   12 21 53 &	5.6    & 39.8	     &    1.25  &  18.5 	&   NGC~4374&
   0.004 &   2.4  &  -12.4      & ... &10  \\
\\
ULX&  12  48 25.9 &   08 30 20 &	...    & 39.5	     &    0.43  &  18.9, 21.9	&   NGC~4698&
   0.003 &   1.2  &  -12.0, -9.0& BL Lac & 11\\
\\
\end{tabular}
{\footnotesize References:
1 This work;~
2 Arp, H. \& Sulentic, J. 1979, ApJ, 229, 496;~
3 F. La Franca, F., Fiore, F., Vignali, C., et al. 2002, ApJ, 570, L100;~	   
4 Arp, H., Burbidge, E. M., Chu, Y., Flesch, E., Patat, F., \& Rupprecht, G. 2002, A\&A, 391, 833;~
5 Arp, H. 1977 ApJ, 218,70;~
6 Masetti, N. et al. 2003, ApJ, 406, L27;~
7 Knezek, P. M., \& Bregman, J. N. 1998, AJ, 115, 1737;-
8 Vogler, A.  \& Pietsch, W. 1999, A\&A, 352, 64;-
9 Stocke, J. T. et al. 1991, ApJS, 76, 813;~
10 Burbidge, G., Hewitt, A., Narlikar, J. V. \&  Das Gupta, P. 1990, ApJS, 74, 675;~
11 Foschini, L., Ho, L., \& Masetti, N. et al. 2002, A\&A, 396, 787}
%\end{center}
%\end{table}

\normalsize
\newpage

\begin{figure}
\begin{center}
\includegraphics[width=10.0cm,angle=-90]{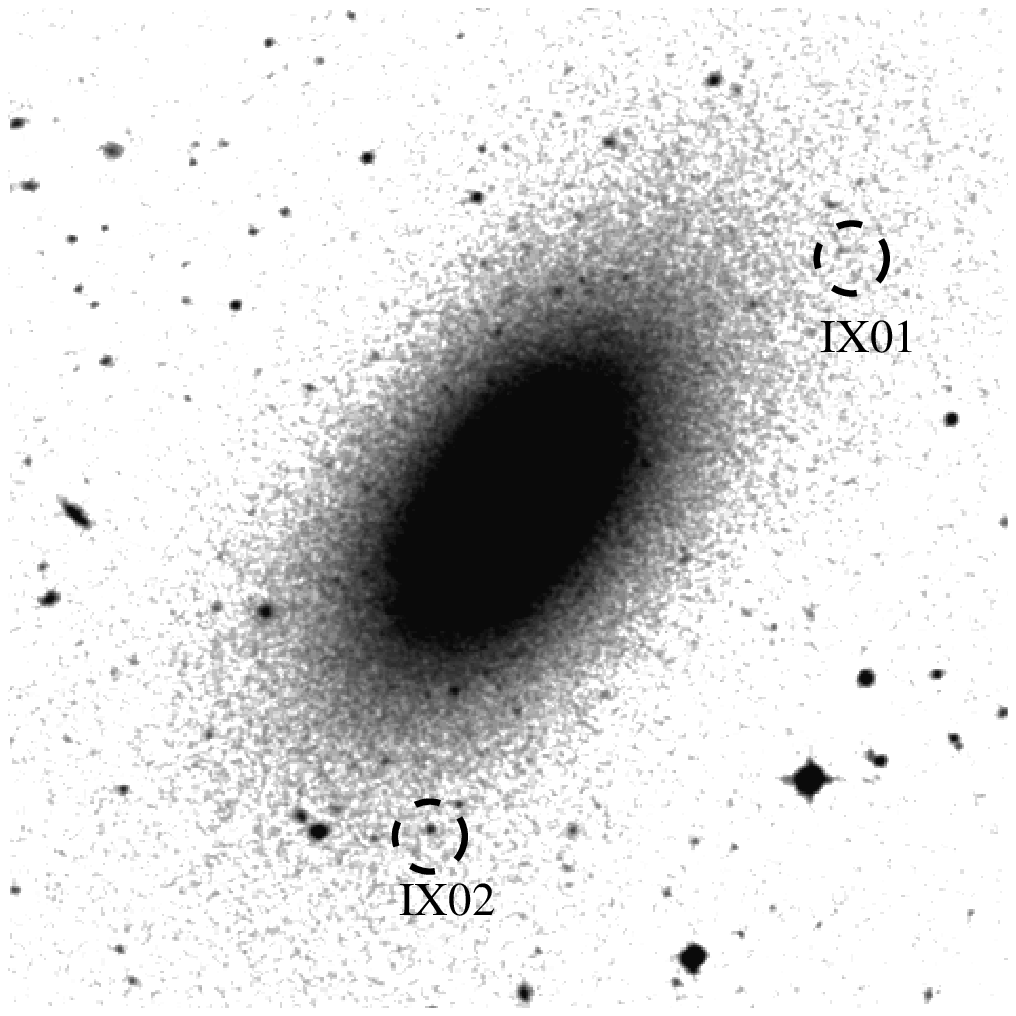}
\includegraphics[width=10.0cm,angle=-90]{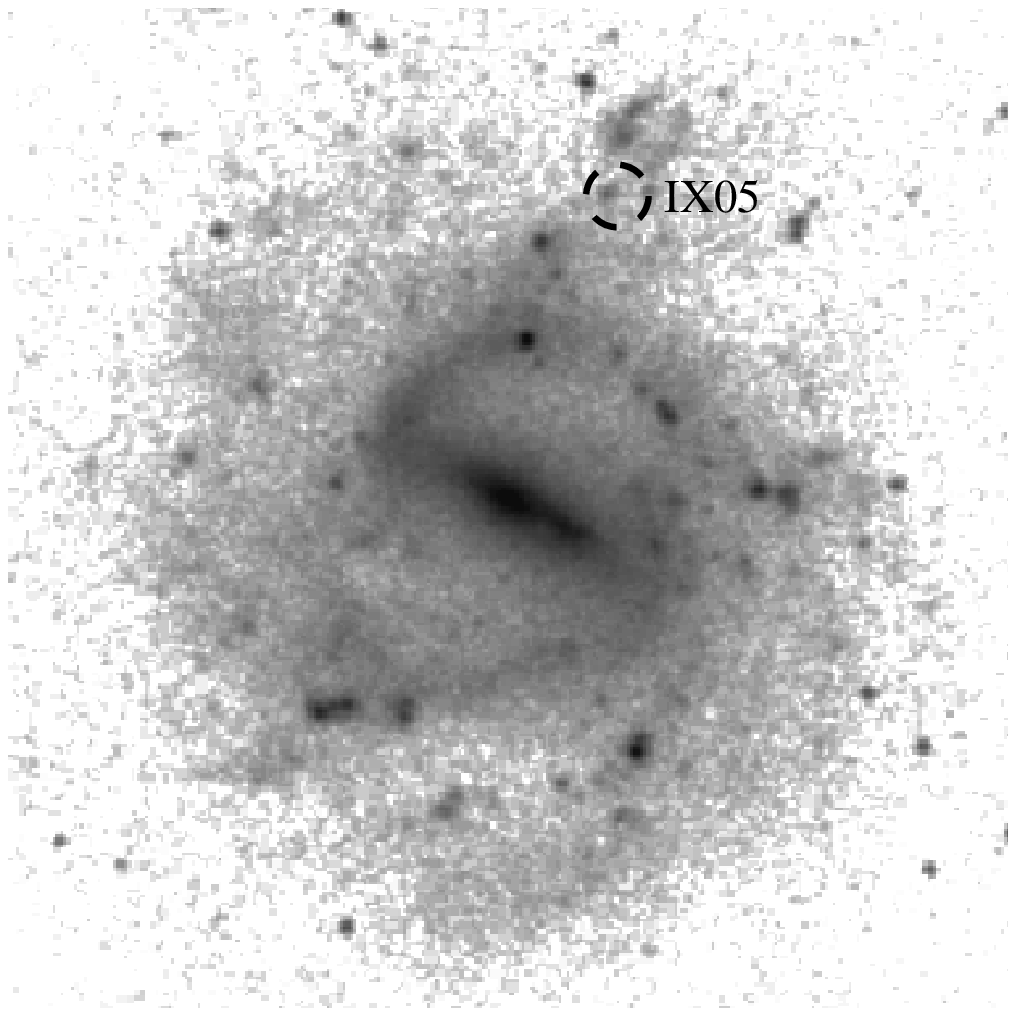}
\caption{DSS plate centred on NGC~720 and NGC~1073 with field of views of $5^{\prime}\times 5^{\prime}$.   
Indicated are the positions of the optical counterparts of the 
objects IXO1, IXO2 and IXO 5.  See main text for the details.}
\label{fig1}
\end{center}
\end{figure}

\begin{figure}
\begin{center}
\includegraphics[width=10.0cm]{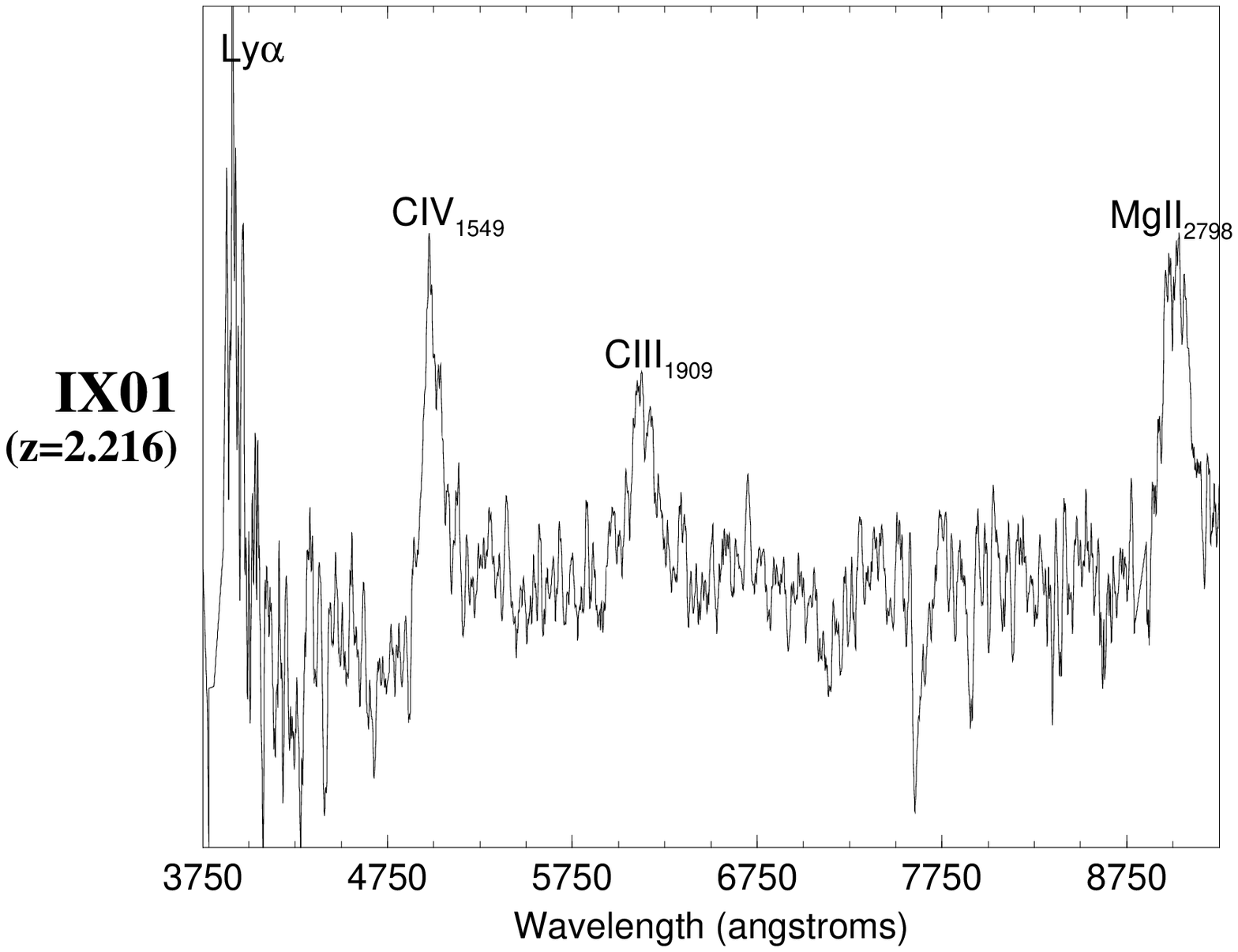}
\includegraphics[width=10.0cm]{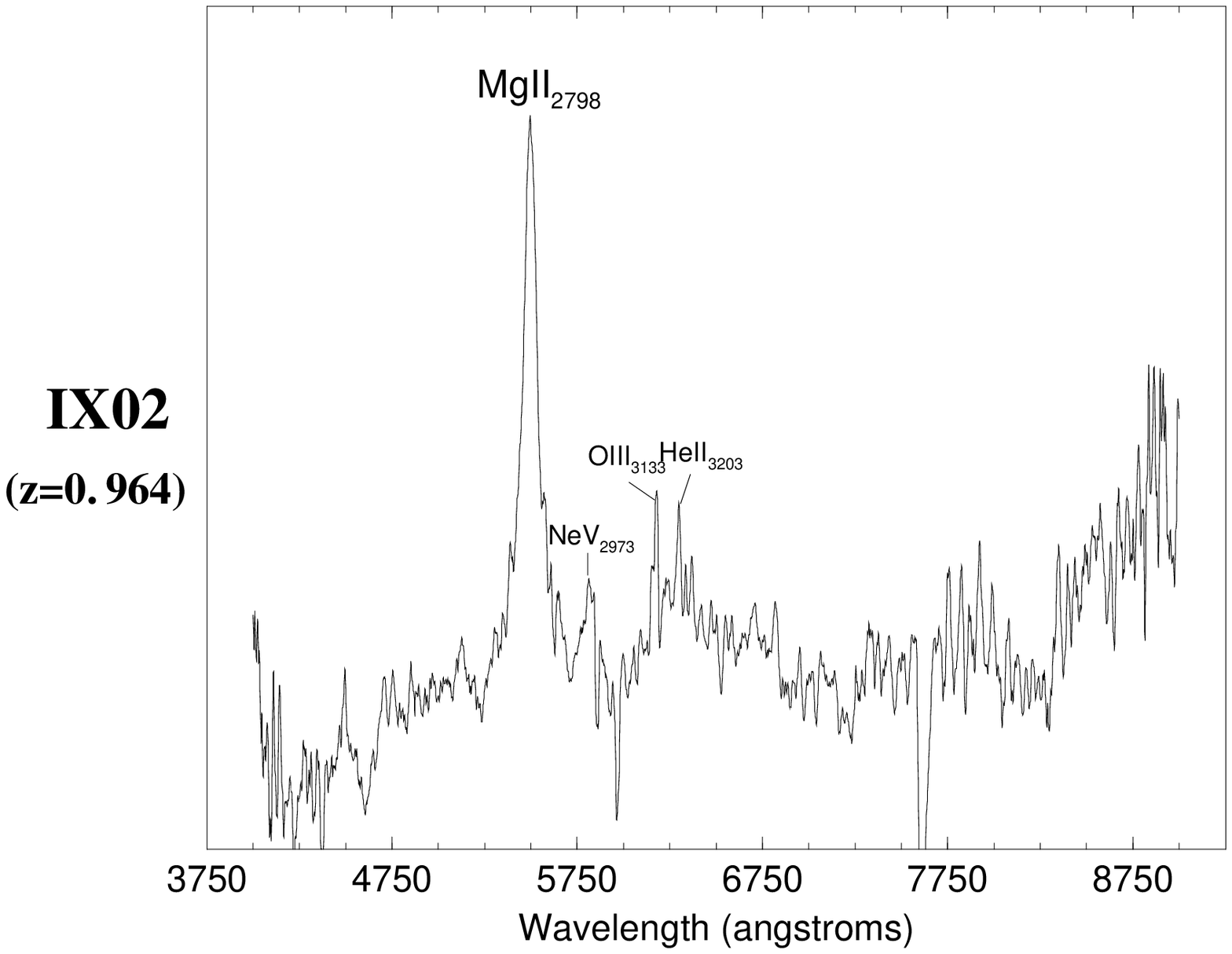}
\includegraphics[width=10.0cm]{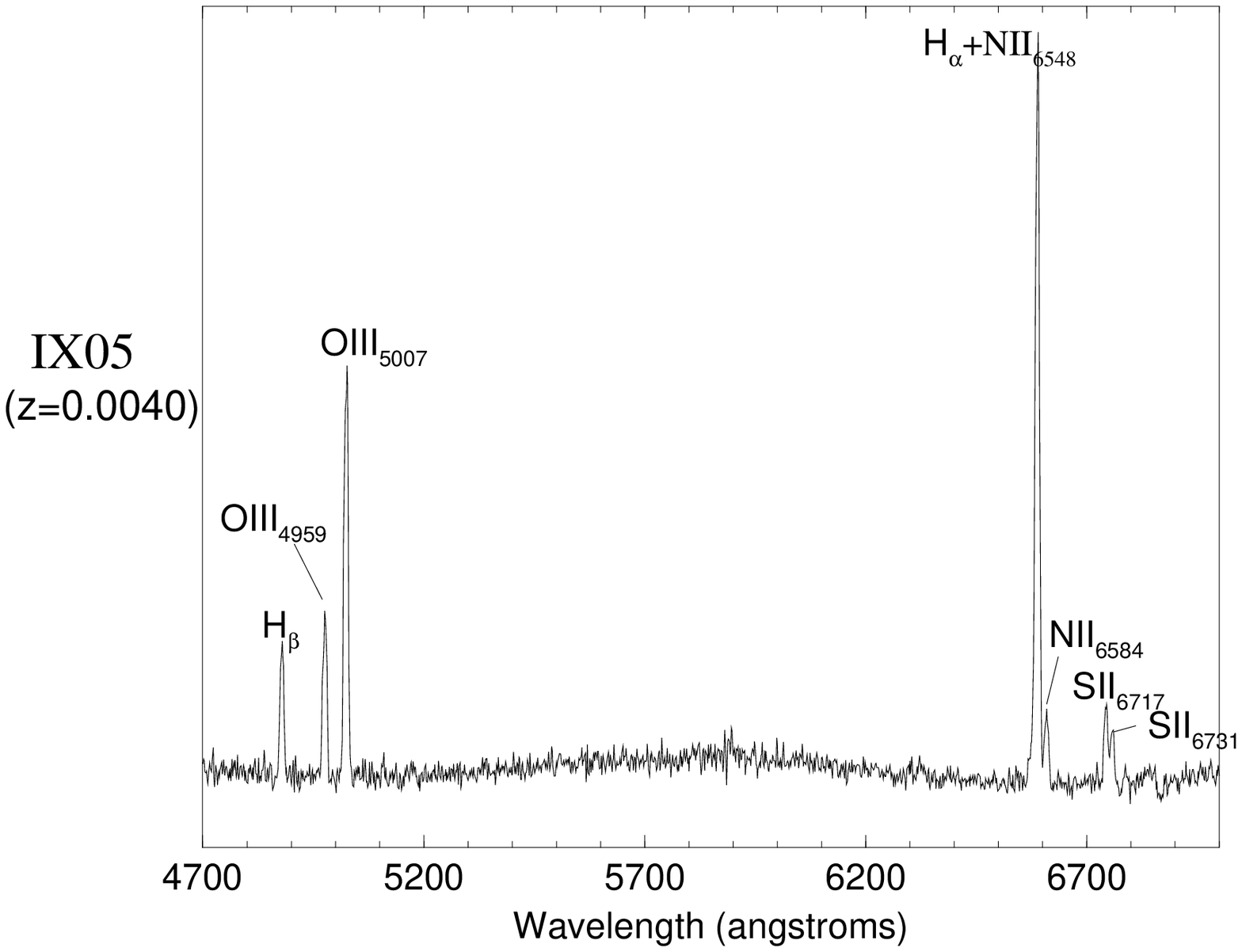}

\caption{Spectra measured with the W. Herschel telescope of the ULX called 
IX01, IX02, IX05 in the Colbert and Ptak Catalog (2002). The vertical axis is the instrumental
flux in arbitrary units.}
\label{fig2}
\end{center}
\end{figure}

%\begin{figure}
%\begin{center}
%\includegraphics[width=10.0cm]{f3b.eps}
%\caption{The relation of optical to X-ray luminosity for the ULX's in Table 2.}
%\label{fig3}
%\end{center}
%\end{figure}

\end{document}